\documentclass[10pt,reprint]{revtex4-1}
\usepackage{graphicx}
\usepackage{hyperref}
\usepackage{amsmath,amssymb,dsfont,mathrsfs,mathtools}
\usepackage{enumerate}

\newcommand{\RR}{\mathds{R}}
\newcommand{\CC}{\mathds{C}}
\newcommand{\EE}{\mathbb{E}}
\newcommand{\sign}{\mathop{\mathrm{sign}}}
\newcommand{\eps}{\varepsilon}

\begin{document}
	
\title{Instantons for rare events in heavy-tailed distributions}

\author{Mnerh \surname{Alqahtani}}
\affiliation{Mathematics Institute, University of Warwick, Coventry
  CV4 7AL, United Kingdom}
\author{Tobias \surname{Grafke}}
\affiliation{Mathematics Institute, University of Warwick, Coventry
  CV4 7AL, United Kingdom}
\date{\today}

\begin{abstract}
  Large deviation theory and instanton calculus for stochastic systems
  are widely used to gain insight into the evolution and probability of
  rare events. At its core lies the realization that rare events are,
  under the right circumstances, dominated by their least unlikely
  realization. Their computation through a saddle-point approximation
  of the path integral for the corresponding stochastic field theory
  then reduces an inefficient stochastic sampling problem into a
  deterministic optimization problem: finding the path of smallest
  action, the instanton. In the presence of heavy tails, though,
  standard algorithms to compute the instanton critically fail to
  converge. The reason for this failure is the divergence of the
  scaled cumulant generating function (CGF) due to a non-convex large
  deviation rate function. We propose a solution to this problem by
  ``convexifying'' the rate function through a nonlinear
  reparametrization of the observable, which allows us to compute
  instantons even in the presence of super-exponential or algebraic
  tail decay. The approach is generalizable to other situations where
  the existence of the CGF is required, such as exponential tilting in
  importance sampling for Monte-Carlo algorithms. We demonstrate the
  proposed formalism by applying it to rare events in several
  stochastic systems with heavy tails, including extreme power spikes
  in fiber optics induced by soliton formation.
\end{abstract}

\maketitle

\section{Introduction}

In many situations of physical relevance, rare events are tremendously
important despite their infrequent occurrence: Heat waves, stock
market crashes, or earth quakes all occur with small probability but
devastating consequences. Unfortunately, due to their rareness, these
events are hard to observe in experiment or numerical simulation, and
require special treatment. Rare event algorithms~\cite{bucklew:2013}
are typically based on one of the two following ideas: Either to
increase the rate of occurrence of the rare event by biasing the
underling system (importance sampling), or to substitute all possible
ways of observing a rare event by its most common realization
(large deviations/instanton theory). Under the hood both are connected to the
\emph{exponentially tilted} measure and the \emph{cumulant generating
  function}. As we will see, when naively implementing standard
schemes, both become ill-defined when the underlying probability
densities become heavy-tailed.

Here, we focus on numerical algorithms connected to instanton theory
and its rigorous counterpart, large deviation theory, to recover
the tails of probability distributions in a stochastic system. A large
deviation principle (LDP) states that the probability of rare events
decays exponentially, and its exponential scaling is given by the
minima of the corresponding rate function
$I$~\cite{varadhan:1966,dembo-zeitouni:2010}. More precisely, let
$P^\eps$ be a family of probability measures on a suitable space
$\mathcal{X}$. We say $P^\eps$ satisfies an LDP with rate function
$I:\Omega\to\RR$ if for all $\Omega \subset \mathcal{X}$, we have
\cite{freidlin-wentzell:2012},
\begin{equation}
  P^\eps\left(\Omega \right) \asymp \exp\left(-\eps^{-1} \,\inf\limits_{x \in \Omega}I\left(x\right) \right),  \label{eq:LDP2}
\end{equation}
where $\asymp$ denotes log-asymptotic equivalence in the limit
$\eps\to0$.

In a physical sense, the probability $P^\eps(\Omega)$ can formally be
written as a path integral, and the estimate~(\ref{eq:LDP2}) becomes
a saddle point approximation or Laplace method. In large deviation theory, the \emph{G\"artner-Ellis theorem}~\cite{ellis-etal:1984}
provides a direct formula for the rate function $I$. Roughly speaking,
if the limiting behavior of a scaled CGF is well-defined, then its
\emph{Legendre-Fenchel} (LF) transform is the rate function of the LDP
of the process under consideration. The LF transform of a real-valued
function $f\left(x\right)$ defined on $\RR^n$ is defined as
\begin{equation}
f^* \left(y\right) = \sup\limits_{x \in \RR^n}
\left(\left\langle x,y\right\rangle - f\left(x\right)\right),
\label{eq:LF trans}
\end{equation}
where $\left\langle x,y\right\rangle = x^T \, y$ is the inner product
on $\RR^n$. Let $z^\eps$ be a sequence of random variables in
$\RR^n$, with probability measures $P^\eps $, and assume
that its scaled CGF, defined as the limit,
\begin{equation}
  \label{eq:cgf}
  G \left(\lambda\right) \equiv \lim\limits_{\eps \to 0} \,
  \eps \log  \, \EE  \left[e^{\eps^{-1}\langle \lambda,
      z^\eps \rangle}\right],
\end{equation}
exists for each $\lambda \in \RR^n$ and is differentiable in
$\lambda$. Then, the G\"artner-Ellis theorem states that the family of
probability measures $\left\{P^\eps \right\}$ satisfy an LDP, where
the rate function $I$ is the LF transform of $G$, $I = G^*$.

Intuitively, one can interpret $\lambda$ as a Lagrange multiplier to
condition on an outcome $z$. Crucially, though, if the probability
measure $P^\eps$ is super-exponential, or has even heavier tails, the
expectation in equation~(\ref{eq:cgf}) diverges and the CGF is no
longer defined. Notably this does not mean that the corresponding rare
events are special in any way, but merely that the duality between the
parameter $\lambda$ and rare event observable $z$ is broken. As a
consequence, no tilt exists to realize an outcome $z$, and standard
rare events algorithms fail.

In what follows, we will show how a \emph{nonlinear tilt} allows to
modify the connection between tilt $\lambda$ and outcome $z$, so that
heavy tails can be probed regardless of the non-convexity of their
rate function. We will concentrate specifically on the case of small
noise sample-path large deviations for stochastic differential
equations (SDEs), which will we introduced in section
\ref{sec:inst-freidl-wentz}.  Here, we focus on the numerical
computation of the instanton in section \ref{sec:instanton_approach},
and highlight the connection to a change of measure in path-space in
\ref{sec:change-prob-meas}. We demonstrate the problem in the
heavy-tailed case by reviewing the convex analysis for the
 G\"artner-Ellis theorem in section~\ref{sec:conv-analys-gartn}, 
 and propose a solution modifying the instanton computation to yield 
 finite outcomes for heavy tails and non-convex rate functions 
 in section~\ref{sec:modif-inst-appr}. To demonstrate the applicability 
 of our approach, we show several examples of instantons for heavy-tailed 
 distributions in section~\ref{sec:applications}: Toy models with 
 super-exponential (section~\ref{sec:stretch-expon-distr}) and powerlaw
(section~\ref{sec:power-law-distr}) tails, and a banana-shaped
potential (section~\ref{sec:banana-potential}), and finally
high-amplitude events in fiber-optics described by the focusing
nonlinear Schr\"odinger equation
(section~\ref{sec:nonl-schr-equat}). We summarize our findings in
section~\ref{sec:conclusion}.
 
\section{Instantons and Freidlin-Wentzell theory}
\label{sec:inst-freidl-wentz}

Consider a stochastic system,
\begin{equation}
  d X^\eps_t = b\left(X^\eps_t\right) \, dt + \sqrt{\eps} 
  \sigma \, d W_t\,,\quad X^\eps_{t_0} = x_0,  \label{eq:smallPerEq}
\end{equation}
where $X^\eps_t\in\RR^n$ is a family of random processes indexed by
the noise strength $\eps$. The deterministic drift $b:\RR^n\to\RR^n$
satisfies the Lipschitz condition, $dW_t$ is an $n$-dimensional
Brownian increment, and the noise covariance $\chi = \sigma \sigma^T$
is assumed to be invertible for $\sigma \in \RR^{n\times
  n}$. Intuitively, equation~(\ref{eq:smallPerEq}) describes the
temporal evolution of a system perturbed by stochasticity, where we
later assume the fluctuations to be small, $\eps\ll1$. This situation
is ubiquitous in many application areas, where for example $\eps$
plays the role of the temperature in a chemical reaction, or the
inverse number of particles in a thermodynamic system.

With vanishing noise, $\eps = 0$, the solution $x\in\RR^n$ of the
unperturbed (deterministic) system
\begin{equation}
  \dot{x}(t) = b\left(x(t)\right), \,\,\,\,  x(t_0) = x_0,   \label{eq:UnPerEq}
\end{equation}
converges to one of its attractors for long times. For example,
consider a point attractor, or \emph{asymptotically stable fixed
  points}, $\bar x\in\RR^n$, with basin of attraction $B$, such that
$x(t)\to\bar x$ for $t\to\infty$ for all initial conditions $x_0$ in
$B$. Solutions of the stochastic system~(\ref{eq:smallPerEq}) converge
to solutions of the deterministic system~(\ref{eq:UnPerEq}) in
probability, $P (\lim\limits_{\eps \rightarrow 0} \, \max\limits_{t_0
  \leq t \leq t_1} \left|X^\eps_t - x(t)\right| = 0) = 1$~
\cite{freidlin-wentzell:2012}. This is an instance of the law of large
numbers, stating that for small noise and large times we expect
solutions of the stochastic system to end up near the attractors of
the deterministic one.

Nevertheless, for any non-zero $\eps \ll 1$ there is a small but non-vanishing 
probability of finding the system far from the attractor. This can
only happen if the noise conspires in just the right way to overcome
the deterministic dynamics, and is consequently a rare
event. Concretely, consider any domain $D \subset\RR^n$ attracted to
$\bar x$, i.e.~$D\subset B$. We are interested in the chance of
trajectories $X_t^\eps$ departing from $\bar{x}$ and eventually
leaving $D$. These trajectories belong to the set
\begin{equation} 
  A_z \coloneqq \left\{\varphi \in \textbf{C}_{t_0 \, t_1}
  \left(\RR^n\right)| \varphi(t_0)= \bar{x}, \varphi(t_1)= z \notin
  D\right\}, \label{eq:set_Az}
\end{equation}
and we want to quantify the probability
\begin{equation}
  \label{eq:P-lownoise}
  p(z) = P\left[X^\eps \in A_z \right]\quad\text{as}\quad\eps\rightarrow 0\,,
\end{equation}             
which is a question about the probability of \emph{large
  deviations}. Under the stated conditions, there is a trajectory
$\varphi^* \in A_z$ such that the probability measure over $A_z$
accumulates near $\varphi^*$ for $\eps\to0$, namely if
$N\left(\varphi^*\right)$ is any neighborhood of $\varphi^*$,
\begin{equation}
  \lim\limits_{\eps \rightarrow 0} \, 
  \frac{P\left[X^\eps \in A_z \setminus N\left(\varphi^*\right)\right]}
       {P \left[ X^\eps \in N\left(\varphi^*\right)\right]}= 0.
\end{equation}
In other words, in the small noise limit we will almost surely find
our sample trajectory close to $\varphi^*$, such that ${\max\limits_{t_0
    \leq t \leq t_1} \left|X^\eps_t-\varphi^*_t\right|\leq \delta}$,
for an arbitrary small $\delta$.

In order to find this most likely trajectory $\varphi^*$,
Freidlin-Wentzell theory~\cite{freidlin-wentzell:2012} states that
$\varphi^*$ is actually the minimizer of large deviation \emph{rate
  function} $S(\varphi)$ associated with the stochastic
system~(\ref{eq:smallPerEq}), given by
\begin{equation}
  S(\varphi) = 
  \frac{1}{2} \int_{t_0}^{t_1} \left\|\dot{\varphi}_t - 
      b\left(\varphi(t)\right) \right\|_\chi^2 dt, \label{eq:FW}
\end{equation}
where the integral exists, and $S(\varphi)=\infty$ otherwise. The norm
$\left\|f\right\|_\chi^2 = \left\langle f, \chi^{-1} f\right\rangle$
is induced by the noise covariance $\chi$. With this rate function we
can quantify the probability~(\ref{eq:P-lownoise}) of departing the domain $D$ as
\begin{equation}
  \lim\limits_{\eps \rightarrow 0} \, \eps \, \log 
  \, p(z)  = -I(z)= 
  - S\left(\varphi^*\right)\,,
\end{equation}
where
\begin{equation}
  \varphi^* = \underset{\varphi \in A_z}{\mathrm{argmin}} \,
  S\left(\varphi\right)\,. \\ \label{eq:argmin}
\end{equation}
The problem of finding the rare event probability is now reduced to
finding the minimizer $\varphi^*$.

In analogy to the principle of least action in classical mechanics or
quantum mechanics, the rate function is often termed \emph{action},
and the corresponding minimizer $\varphi^*$ is called
\emph{instanton}. The integrand of $S$ can be understood as a
\emph{Lagrangian},
\begin{equation}
  L \left(\varphi, \dot{\varphi}\right) = \frac{1}{2} 
  \left\|\dot{\varphi}_t - b\left(\varphi\right) \right\|_\chi^2,
   \label{eq:LFW}	
\end{equation}
so that the maximum likelihood pathways leaving the attractors
of~(\ref{eq:smallPerEq}) correspond to semi-classical trajectories of
the field theory defined by $L$.

\subsection{Instanton equations and large deviation Hamiltonian}
\label{sec:instanton_approach}

It is helpful, both to increase understanding, and to simplify the
numerical implementation, to rephrase the optimization
problem~(\ref{eq:argmin}) into the corresponding Hamiltonian
formulation. To this end, we introduce the large deviation
\emph{Hamiltonian} $H(\varphi, \vartheta)$ as the Legendre transform of
the Lagrangian $L\left( \varphi, \dot{\varphi} \right)$,
\begin{equation}
  H\left(\varphi, \vartheta \right) = \sup\limits_{\dot{\varphi}} \left( \left\langle  
  \vartheta, \,\dot{\varphi} \right\rangle  - \textit{L}\left(\varphi, \dot{\varphi} \right)\right)\,,
    \label{eq:def-p} 
\end{equation}
which, for the Lagrangian~(\ref{eq:LFW}), corresponds to
\begin{equation}
  H(\varphi,\vartheta) = \langle b(\varphi), \vartheta\rangle + \tfrac12 \langle \vartheta, \chi \vartheta\rangle\,.
\end{equation}
Here $\vartheta = \partial L/\partial \dot{\varphi}$ is the conjugate
momentum of $\varphi$ \cite{deriglazov:2017}. Now, the minimizer
$\varphi^*$ can also be expressed as the solution of
Hamilton's equations,
\begin{equation}
  \begin{split}
    \dot{\varphi} & =  \partial_\vartheta H(\varphi,\vartheta) = b\left(\varphi\right) + \chi \ \vartheta, \\
    \dot{\vartheta} & =   -\partial_\varphi H(\varphi,\vartheta) = - \left(\nabla_{\varphi} b\left(\varphi\right)\right)^T \, \vartheta.\label{eq:varphi-p}
  \end{split}
\end{equation}
with boundary conditions, 
\begin{equation}
\varphi(t_0) = \bar x, \qquad   \varphi(t_1)=z. \label{eq:boundary_con}
\end{equation}
Equations~(\ref{eq:varphi-p}) and~(\ref{eq:boundary_con}) are often
termed \emph{instanton equations}.

The fact that we are looking only for trajectories that will
eventually leave the attractor, $\varphi^*\in A_z$, implies that the
optimization problem (\ref{eq:argmin}) is a constrained
one, i.e.~we are looking only for solutions of the instanton equations
conditioned on the endpoint $z$. Practically, this
constrained optimization problem can be transformed into an
unconstrained one,
\begin{equation}
  \varphi^* =  \underset{\varphi \in \textbf{C}_{t_0 \, t_1}\left(\RR^n\right)}{\mathrm{argmin}} \left(S\left(\varphi\right)
  - \left\langle \lambda, \ \varphi(t_1) - z \right\rangle \right),  \label{eq:Lagrange}  
\end{equation}
by using a Lagrange multiplier $\lambda\in\RR^n$ to enforce the final
constraint~\cite{rindler:2018}. Note that the variation of this
unconstrained action,
\begin{equation}
  \label{eq:unconstrained}
  \left[ \delta S\left(\varphi\right) - \left\langle \lambda, \, 
    \delta \varphi(t_1)\right\rangle  \right]_{\varphi = \varphi^*} = 0,
\end{equation}
results in the same instanton equations,(\ref{eq:varphi-p}), but with
different boundary conditions
\begin{equation}
  \varphi\left(t_0\right)=0,\quad\vartheta\left(t_1\right)
  =  \lambda.  \label{eq:conds}
\end{equation}
We can solve the instanton equations~(\ref{eq:varphi-p}) iteratively
with these conditions, by solving the $\varphi$-equation forward in
time, and using the result to solve the $\vartheta$-equation backward
in time, until convergence~\cite{chernykh-stepanov:2001,
  grafke-grauer-schaefer-etal:2014}. Note that this choice of temporal
direction of integration is not only the one suggested by the boundary
conditions, but is further the numerically stable choice of direction
for the drifts $b$ and $\nabla b^T$.

As we will see below, the mapping between Lagrange multipliers
$\lambda$ and final points $z$ is nontrivial, and it is not clear
\emph{a priori} how to choose the correct $\lambda$ to obtain a final
configuration $z = \varphi^*(t_1)$. If we are interested in
$p^\eps(z)$ for a whole range of $z$, we can instead choose to simply
solve the instanton equations for a range of $\lambda$ to cover a
range of $z$ without specifically needing to know the duality mapping
$\lambda(z)$. Exactly this procedure is often used in the literature
to work out probability distributions of stochastic systems, from
Burgers~\cite{chernykh-stepanov:2001, grafke-grauer-schaefer:2013} or
Ginzburg-Landau~\cite{rolland-bouchet-simonnet:2016} equations to the
Kardar-Parisi-Zhang~\cite{meerson-katzav-vilenkin:2016} and
Kipnis-Marchioro-Presutti model~\cite{zarfaty-meerson:2016}

Crucially, however, the existence of a corresponding dual $\lambda$
for a given final point $z = \varphi^*\left(t_1\right)$ is not
necessarily guaranteed, as the next section clarifies. As a
consequence, the above methodology might fail, in particular in
situations with heavy tails.

\subsection{Exponentially tilted measures}
\label{sec:change-prob-meas}

Interestingly, there is a probabilistic interpretation of the
introduction of the Lagrange multiplier $\lambda$ to the optimization
problem~(\ref{eq:unconstrained}) in the form of the
\emph{exponentially tilted measure}~\cite{touchette:2009,
  cohen-elliott:2015}, as for example employed in importance sampling
for Monte-Carlo estimators~\cite{bucklew:2013}. Intuitively, by the
procedure of \emph{tilting}, one replaces the original random
process~(\ref{eq:smallPerEq}) by a modified one, under which the rare
events under consideration become more likely, while correcting for
this modification \emph{a posteriori} when computing their
probability. As a consequence, with tilting, the rare event
probability, or expectations over its realizations, can be estimated
more efficiently, and with possibly smaller variance.

To be more precise, we denote by $p_\lambda$ the measure exponentially
tilted towards the outcome $z$, defined for our purposes as
\begin{equation}
  p_\lambda(z) = \frac{\exp\left( \eps^{-1}\left\langle 
    \lambda, \, z\right\rangle  \right) }{\mathop{\EE}_p
    \left[\exp\left( \eps^{-1}\left\langle \lambda, \,
      z\right\rangle \right)\right]} \, p\left(z \right),   \label{eq:P_lambd}
\end{equation}
where $\mathop{\EE}_p\left[.\right]$ is the expectation under the
original measure $p$. In equation~(\ref{eq:P_lambd}), the probability
measure $p_\lambda $ of events resulting in $z$, i.e.~trajectories
in $A_z$, have been awarded extra weight by the \emph{Radon-Nikodym
  derivative} of $p_\lambda$ with respect to $p$, which is:
\begin{equation}
  \frac{p_\lambda \left(z \right)}{p \left(z \right)} = 
  \frac{dp_\lambda \left(z \right)}{dp \left(z \right)} = \frac{\exp
    \left( \eps^{-1}\left\langle \lambda, \, z\right\rangle  \right) }
       {\mathop{\EE}_p \left[\exp\left( \eps^{-1}\left\langle \lambda, \, z\right\rangle  \right)\right]}.   
\end{equation}
Rearranging equation (\ref{eq:P_lambd}) gives:
\begin{equation}
  p_\lambda \left(z \right) = \exp\left( \eps^{-1} \,
  \left\lbrace \left\langle \lambda, \, z\right\rangle  - G \left( \lambda \right)
  \right\rbrace \right) \, p\left(z \right), \label{eq:P_lambda2} 
\end{equation}
where
\begin{equation}
  G \left( \lambda \right) = \eps \, \log \, \EE_p
  \left[\exp\left(\eps^{-1} \left\langle \lambda, \,
    z\right\rangle \right)\right]
\end{equation}
is the scaled CGF, $G:\RR^n\to\RR$. This change of measure is optimal,
in the sense of maximizing the tilted probability $p_\lambda$, at the
choice $\lambda = \lambda_z$ with
\begin{equation}
  \lambda_z = \underset{\lambda \in \RR^n}{\mathrm{argmax}} \left\lbrace \left\langle
  \lambda, \, z\right\rangle  - G \left( \lambda \right)  \right\rbrace.
\end{equation}
This can be seen, given the G\"artner-Ellis theorem, $I(z) =
G(\lambda)^* = \sup_{\lambda\in\RR^n}(\langle \lambda,z\rangle -
G(\lambda))$, by realizing that
\begin{equation}
  \begin{split}
    \log p_\lambda \left(z \right) & =  \eps^{-1} \, \left(
    \left\langle \lambda_z, \, z\right\rangle  - G \left( \lambda_z \right) \right)  \,
    + \, \log p\left(z\right) \\
    & =  \eps^{-1} \, \underset{\lambda \in \RR^n}{\mathrm{sup}} \left\lbrace
    \left\langle \lambda, \, z\right\rangle  - G \left( \lambda \right)\right\rbrace   \, + \,
    \log p\left(z \right) \\
    & =  \eps^{-1} \, I\left( z \right) \, + \, \log p\left(z\right) \\
    & \stackrel{\eps\to0}{\longrightarrow} 0, \label{eq:proof_opt_lambda}
  \end{split}
\end{equation} 
where the last line is just the definition of the LDP, $\eps \log p(z)
= -I(z)$ for $\eps\to0$.

It is in this sense that the optimal tilting parameter of the
end-point distribution corresponds to the Lagrange multiplier in the
instanton equations constraining the endpoint to $z$.

\section{Convex analysis and the G\"artner-Ellis theorem}
\label{sec:conv-analys-gartn}

In order to use the described methodology to find the instanton for a
rare outcome $z$, or equivalently make sense of the corresponding
exponentially tilted measure $p_\lambda(z)$, we must demand that the
mapping $z\to\lambda(z)$ is a bijection: For every outcome $z$ there
must be a unique tilt $\lambda$ such that the instanton $\varphi$,
solution of (\ref{eq:varphi-p}) with boundary
conditions~(\ref{eq:conds}) have a unique solution with
$\varphi(t_1)=z$. If that is the case then we can estimate the
probability
\begin{equation}
  \label{eq:pz-LDT}
  p(z) \asymp \exp (-\eps^{-1} S(\varphi^*)) = \exp(-\eps^{-1} I(z))\,.
\end{equation}
The precise properties of the duality mapping between
tilting parameter $\lambda$ and outcome $z$ can be understood by the
interplay between the G\"artner-Ellis theorem and convex analysis. We have,
\begin{equation}
  \begin{split}
    G\left( \lambda \right) & = \sup\limits_{z \in \RR^n}
    \left( \left\langle \lambda, \,\, z \right\rangle- I\left(z\right)\right) \\
    & = \left\langle \lambda, \,\, z\left(\lambda\right)  \right\rangle-
    I\left(z\left(\lambda\right)\right),   \label{eq:Glambda}
  \end{split}
\end{equation}
where the solution $z\left(\lambda\right)$ of the form
\begin{equation}
  \label{eq:nabla_I}
  \nabla I\left(z\right) = \lambda\, ,
\end{equation}
does only hold when the rate function is strictly convex. If instead
the rate function is not strictly convex (i.e.~has concave, and/or
affine linear regions or is even just asymptotically linear), the LF
transform is applied only to the region at which $I\left(z\right)$
admits \emph{supporting hyperplanes}.  If there exists $\lambda \in
\RR^n$ such that \cite{touchette:2005},
\begin{equation}
  \label{eq:supp_hyp}
I\left(y\right) \geq I\left(z\right) + \lambda \left( y- z \right), \ \ \forall   y \in \RR^n,
\end{equation}
then we say $I$ admits a supporting hyperplane at $z$, where the slope
of the supporting hyperplane is $\lambda$. In this sense, we can
define non-convex regions to be the ones that do not admit any
supporting hyperplane, so do not have any corresponding
$\lambda$. Note that the absence of these hyperplanes can affect the LF
transform $I^*\left(z\right) = G \left(\lambda\right)$ in two
different ways,
\begin{enumerate}[\bfseries {Case} I:]
	\item Having an asymptotically linear part of $I\left(z\right)$ leads to a 
	divergent LF transform $G \left(\lambda\right)$.  \label{Case1} 
	\item Having a concave or affine linear part of $I\left(z\right)$ leads to an existent 
	but nondifferentiable LF transform $G \left(\lambda\right)$. \label{Case2}  
\end{enumerate} 
Both cases will be discussed specifically in the applications in
section~\ref{sec:applications}.

Assuming for now there are supporting hyperplanes (i.e.~existent $\lambda$) 
for all $z \in \RR^n$, then equation (\ref{eq:nabla_I}) leads to,  
\begin{equation}
  z\left(\lambda\right) = \left( \nabla I\right) ^{-1}\left(\lambda\right), 
\end{equation}
i.e.~$\nabla I$ must be invertible for $z(\lambda)$ to be so. Up to a
choice of sign, this implies that $\nabla I$ is \emph{strictly
  monotonically increasing} (SMI), which is equivalent to $I$ being a
strictly convex function~\cite{touchette:2009}. Also note that if
$z\left(\lambda \right)$ is invertible, this implies that
$G\left(\lambda\right)$ is a differentiable function, since equation
(\ref{eq:Glambda}) gives,
\begin{equation} 
  \begin{split}
    \nabla G \left( \lambda \right) & = z\left( \lambda \right) \, +
        \left( \nabla z\left( \lambda \right)\right)^T  \,\, \lambda
    -  \left( \nabla z\left( \lambda \right)\right)^T   \nabla I\left(z\left( \lambda \right)\right)  \\
    & = z\left(\lambda\right), \label{eq:G_der}
  \end{split}
\end{equation}
where equation (\ref{eq:nabla_I}) is used. What we have demonstrated
above is nothing but the well-known fact that the LF transform of a
convex, differentiable function $G(\lambda)$ is strictly convex. This
perspective, though, makes it clear that the existence of a tilting
parameter (Lagrange multiplier) $\lambda$ to enforce an outcome $z$
depends on the finiteness and differentiability of the scaled CGF.  In
other words, both exponential tilting, and finding a Lagrange
parameter to constrain the endpoint to $z$, depends on the rate
function being strictly convex.

Since in the low noise limit we have $p(z)\sim\exp(-\eps^{-1} I(z))$, it is
easy to construct cases where the rate function $I(z)$ is not
strictly convex. In fact, every situation where the tails of $p(z)$
are \emph{fat}, i.e.~exponential ($I(z)\sim z$), stretched exponential
($I(z)\sim z^\alpha, \alpha<1$), or even algebraic ($I(z)\sim\alpha
\log(z), \alpha<0$) tails will break the above assumption. Examples of
fat tailed distributions are ubiquitous in physical systems of
relevance, such as the energy dissipation in fluid turbulence and the
phenomenon of intermittency~\cite{frisch:1995}, wealth distributions in
economies~\cite{dragulescu-yakovenko:2001, sinha:2006} or stock price 
changes in finance~\cite{gopikrishnan-plerou-gabaix-etal:2000,
plerou-gopikrishnan-amaral-etal:2000}.

The main contribution of this paper is the realization that the
introduction of a nonlinear map $F:\RR^n\to\RR^n$ allows us to loosen
the restriction of the convexity of $I(z)$. The idea is to define a
\emph{nonlinear tilt} through $F$ via $\exp(\eps^{-1}\langle \lambda,
F(z)\rangle)$, such that the map $\lambda\to z(\lambda)$ is replaced by
a new map $\lambda \to F\circ z(\lambda)$. We are free to choose any
appropriate $F$. As we will see next, this allows us to suitably
reparametrize the space of outcomes, so that the \emph{effective} rate
function $I\circ F^{-1}$ is strictly convex.
 
\subsection{Nonlinear tilt}
\label{sec:modif-inst-appr}

In analogy to equation~(\ref{eq:P_lambda2}) and the description in
sections~\ref{sec:change-prob-meas} and~\ref{sec:conv-analys-gartn},
we can now define the \emph{nonlinearly tilted measure}
\begin{equation}
  \begin{split}
    p_\lambda^F(z) & = \frac{\exp(\eps^{-1}\langle \lambda, F(z)\rangle)}{\EE_p \exp(\eps^{-1}\langle \lambda, F(z)\rangle)} p(z)\\
    & = \exp\left(\eps^{-1} (\langle \lambda, F(z)\rangle -
  G_F(\lambda)) \right) \, p(z)
  \end{split}
\end{equation}
where the nonlinearly tilted CGF is given by 
\begin{equation} 
  \begin{split}
    G_F \left( \lambda \right) & =  \sup\limits_{z \in \RR^n}
    \left( \left\langle \lambda, \, F\left(z\right) \right\rangle 
    - I\left(z\right) \right)  \\ \label{eq:G_F}
    & =   \left\langle \lambda, \, F\left(z\left(\lambda\right)\right)
    \right\rangle  - I\left(z\left(\lambda\right)\right), 
  \end{split}
\end{equation} 
(compare equation (\ref{eq:Glambda})) which is assumed to be finite and 
differentiable. Its gradient fulfills
\begin{equation} 
\begin{split}
\nabla G_F \left( \lambda \right) & = F\left(z\left(\lambda\right)\right) +
   \left(\nabla z(\lambda) \right)^T  \left( \nabla F(z(\lambda))\right)^T  \, \, \lambda \\
& -   \left(\nabla z(\lambda)\right)^T \nabla I(z(\lambda))    \\
& = F\left(z \left(\lambda \right) \right), \label{eq:G_F_der}
\end{split}
\end{equation} 
where the last equality is due to $z\left(\lambda\right)$ being the
solution of $\nabla I\left(z\right) = \lambda^T \nabla F(z)$ in equation
(\ref{eq:G_F}).

This proposed remapping can be chosen to overcome the above problem by
creating a new function $G_F\left( \lambda \right)$, which plays the
role of the CGF, while simultaneously being a bounded and
differentiable function. At the same time, $I \circ F^{-1}
\left(y\right)$ can be understood as the effective rate function,
since equation~(\ref{eq:G_F}) can be written as,
\begin{equation}
G_F \left( \lambda \right) = \sup\limits_{F^{-1} \left(y\right) \in \RR^n}
 \left( \left\langle \lambda, \, y \right\rangle - I \circ F^{-1} \left(y\right) \right).
\end{equation}
Obviously, the right choice of $F$ depends on the nature of the tail
scaling at hand. We will derive the necessary properties of
$F\left(.\right)$ next.
 
\subsection{Properties of the reparametrization and the nonlinearly tilted instanton}
\label{sec:prop-chos-observ}

In the following, we denote by $y\in\RR^n$ the re\-pa\-ra\-me\-trized
outcome, $y=F(z)$. Our goal is to choose $F$ such that $F\circ
z(\lambda) = y(\lambda)$ is a bijection. From above, it is clear that
\begin{equation}
  \lambda^T = \nabla I(z) (\nabla F(z))^{-1}\,.
\end{equation}
Following the same argument as in section~\ref{sec:conv-analys-gartn},
$y(\lambda)$ is bijective if
\begin{itemize}
\item $F$ is a diffeomorphism, and
\item $I\circ F^{-1}(y)$ is strictly convex, i.e.
  \begin{equation}
    \label{eq:gen-convex}
    \langle v, \mathrm{Hess}(I\circ F^{-1})(y)\, v\rangle > 0\ \forall\ v\in\RR^n\,.
  \end{equation}
\end{itemize}
Assuming these conditions on $F$ implies that the gradient of the
reparametrized rate function ${I \circ F^{-1} (y)}$, given by $\lambda
\circ F^{-1} \left(y\right)$, is an SMI function, implying that it is
invertible. The desired bijective mapping then becomes ${\lambda
  \rightarrow F \left(z\left(\lambda\right)\right) = \nabla G_F \left(
  \lambda \right)}$ (compare equation (\ref{eq:G_F_der})). For a
non-convex $I$, intuitively, $F$ must be chosen to suitably
reparametrize the space of outcomes for the effective rate function to
become strictly convex. For example for the $n=1$ case with heavy tails, one
might imagine a strong enough \emph{compression} of the observable $z$
such that a fat tail becomes non-fat.

To harness this nonlinear tilt in the computation of instantons for
distributions with heavy tails, we need to modify the approach outlined
in section~\ref{sec:instanton_approach} as follows: The variation of
the unconstrained action~(\ref{eq:unconstrained}) now reads
\begin{equation}
  \label{eq:unconstrained-tilted}
  \left[ \delta S\left(\varphi\right) - \left\langle \lambda, \, 
    \nabla F(\varphi(t_1)) \delta \varphi(t_1)\right\rangle  \right]_{\varphi = \varphi^*} = 0\,.
\end{equation}
Consequently, the boundary conditions of the instanton equations are
modified to 
\begin{equation}
  \label{eq:bnd-cond-tilted}
  \varphi\left(t_0\right)=0,\quad\vartheta\left(t_1\right)
  =  \lambda \nabla F(\varphi(t_1))\,,
\end{equation}
which will yield an instanton trajectory $\varphi^*$ that reaches $z$,
$\varphi(t_1)=z$, despite the fact that the rate function $I(z) =
S(\varphi^*)$ is not convex around $z$. Since $F$ is continuous, the
probability measure $P^\eps \circ F^{-1} \left(y\right)$ in the limit
$\eps\to0$ is the same as $P^\eps \left(z\right)$ for a continuous
$F$, according to the contraction
principle~\cite{freidlin-wentzell:2012, dembo-zeitouni:2010}.

Note that this reparametrization through $F$ is introduced solely to
adequately define the tilted measure, or equivalently numerically
compute the instanton without encountering divergences. Afterwards,
the reparametrization can be reverted to obtain the probability
distribution in the original coordinates $z$. 

As additional remark, methods that compute the instanton by solving
the global optimization problem, for example by solving the associated
Euler-Lagrange equations instead of integrating the instanton
equations~\cite{e-ren-vanden-eijnden:2004,
  grafke-schaefer-vanden-eijnden:2017}, do not require the above
treatment: The tilting parameter disappears in these cases as the
boundary conditions are fixed in the field variable instead of the
conjugate momentum. Therefore, in principle, these methods can be
chosen in the non-convex case. The solution of the instanton
equations, though, is generally
preferred~\cite{grafke-vanden-eijnden:2019} due to numerically
efficiency, and sometimes even required (such as when the noise
covariance in~(\ref{eq:smallPerEq}) is not invertible).

\section{Applications} \label{sec:applications}

We will now consider a number of examples that show how to compute
tail probabilities in stochastic systems. To demonstrate the wide
applicability of our approach, we consider several cases that
highlight different complications. We start with two toy models that
feature stretched exponential (section~\ref{sec:stretch-expon-distr})
or powerlaw (section~\ref{sec:power-law-distr}) tails. Then, in
section~\ref{sec:banana-potential}, we consider a two-dimensional
system with a bent (``banana-shape'') potential, where the
non-convexity is not due to heavy tails, but due to the shape of the
unimodal invariant probability density. Lastly, we demonstrate the
practical applicability of our method by considering an example
motivated from fiber optics in section~\ref{sec:nonl-schr-equat}. Here
we compute the probability of measuring extreme power spikes at the
end of extended optical fibers, where the probability distribution of
the input signal is known. Due to soliton formation, this distribution
features heavy tails for long fiber lengths ($L\gg 10\ \textrm{m}$),
so in order to compute probabilities via an instanton approach, our
corrections are necessary.

\subsection{Stretched exponential}
\label{sec:stretch-expon-distr}

Consider the stochastic gradient flow,
\begin{equation}
  \label{eq:gradient_flow}
  d X^\eps_t = -\nabla U(X^\eps_t) \, dt + \sqrt{2 \,\eps} \, \, d W_t\,,\quad t\in[t_0,t_1]\,.
\end{equation}
The potential $U: \RR^n \rightarrow \RR$ determines completely the
stationary probability distribution function (PDF)
\begin{equation}
  \label{eq:F_P_rho}
  \rho_\infty(z) = Z^{-1} \exp \left( - \eps^{-1}  U\left(z\right) \right)
\end{equation}
with normalization constant $Z$. We further assume that $U$ has a
unique minimum, i.e.~we are only considering unimodal
distributions. For large times, $t_1-t_0 = T\to\infty$, the
distribution of endpoints of $X^\eps_{t_1}=z$ will converge to
$\rho_\infty(z)$. From the perspective of large deviation theory
(LDT), comparing~(\ref{eq:pz-LDT}) to~(\ref{eq:F_P_rho}), the rate
function for the final point distribution is equivalent to the
potential, $I(z) = U(z)$, and
\begin{align}
  \lim\limits_{\eps \rightarrow 0 } \eps \, \log \, p(z)
  = & \, - U\left(z\right)\,.
\end{align}
Therefore, in order to approximate the tails of the stationary
distribution, we can compute the instanton $\varphi^*$ ending at $z$
and estimate $\rho_\infty(z) \approx \exp(-\eps^{-1} S(\varphi^*))$.

\begin{figure}
  \begin{center}
    \includegraphics[width=246pt]{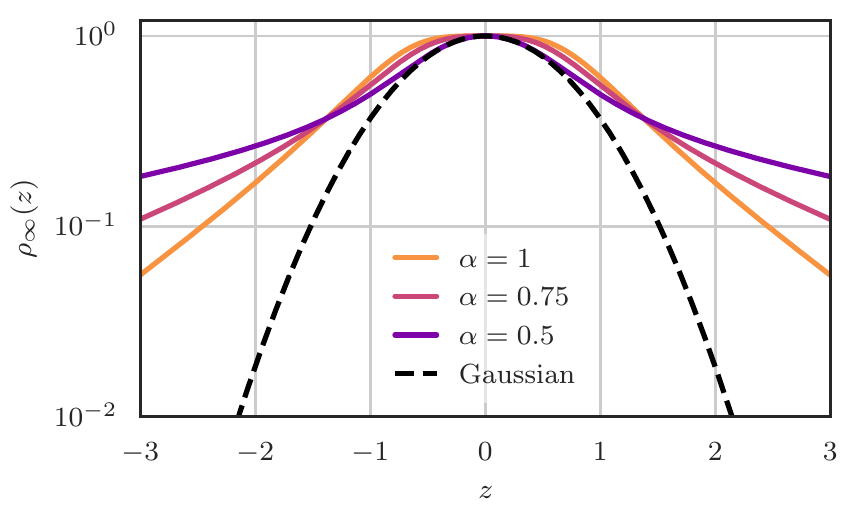}
  \end{center}
  \caption{The potential ${U\left( z \right) = \left(z^4/(1 +
      \left|z\right|^3\right)^\alpha}$, ${0 < \alpha \leq 1}$, for the
    gradient flow SDE (\ref{eq:gradient_flow}) leads to heavy
    (stretched exponential) tails of the stationary density
    $\rho_\infty(z)$ as $\alpha$ decreases.  \label{fig:U_phi_diff}}
\end{figure}

We choose $n=1$ and consider the non-convex potential,
\begin{equation}
  \label{eq:ratefct}
  U\left( z \right) = \left(\frac{z^4}{1 + \left|z\right|^3 }\right)^\alpha\,,\quad 0 < \alpha \leq 1\,,
\end{equation}
which corresponds to a stretched exponential stationary distribution:
At the tails, the dominant exponent is $\alpha$, and $\rho_\infty(z)
\approx \exp(-\eps^{-1}\left|z\right|^{\alpha})$ for large $z$, as
shown in figure~\ref{fig:U_phi_diff}. For this distribution,
$\EE[\exp(\eps^{-1}\lambda z)]$ diverges for large $\lambda$ as in case~\ref{Case1}, 
and hence numerical methods to find the instanton fail in the tail.

\begin{figure}
  \begin{center}
    \includegraphics[width=246pt]{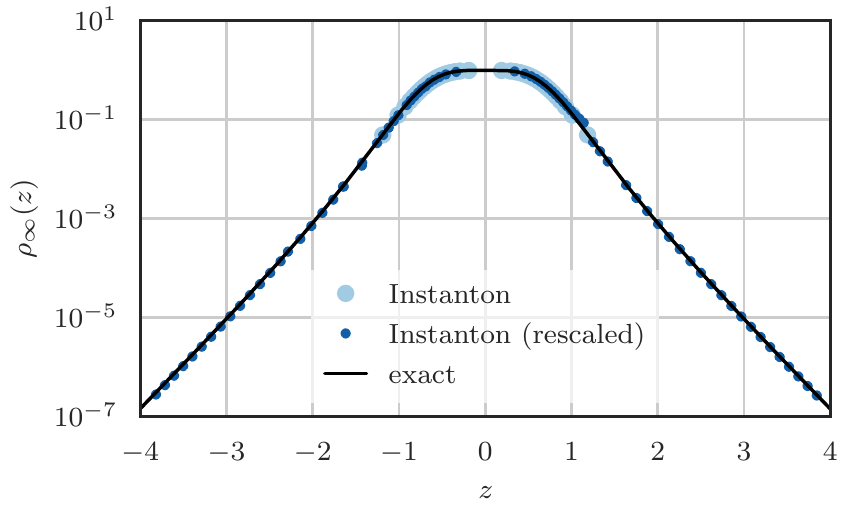}
  \end{center}
  \caption{Stationary density with exponential tails
    (equation~(\ref{eq:ratefct}) for $\alpha=1$). Probing the tails
    with the traditional instanton method (light blue) leads to
    numerical divergence around the non-convex tail
    region. Reparametrizing the observable via $F(z) = \sign(z) \log
    \left| z\right| $ convexifies the tail, so that the instanton (dark
    blue) correctly predicts
    the exact tail probabilities (solid black). \label{fig:stretched-exponential-pdf}}
\end{figure}

\begin{figure}
  \begin{center}
    \includegraphics[width=246pt]{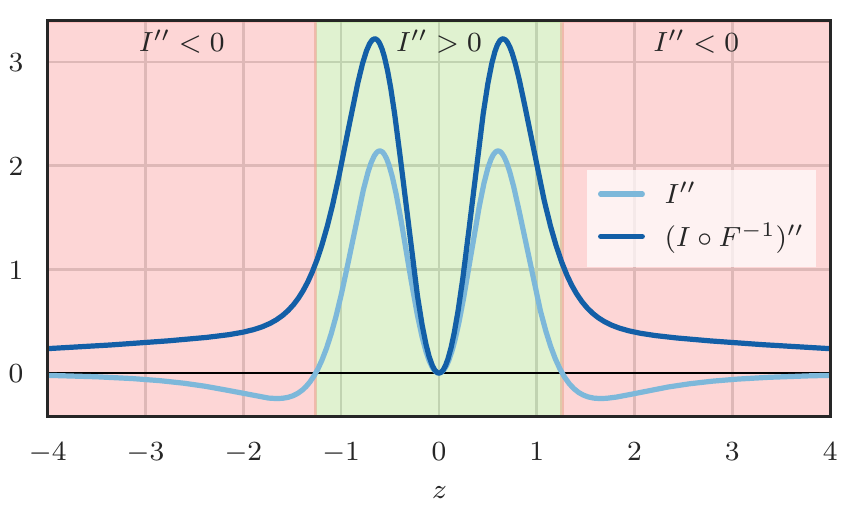}
  \end{center}
  \caption{Convexity condition for the stretched exponential tails:
    The second derivative of the rate function becomes negative beyond
    the inflection points at $z\approx \pm 1.26$. The nonlinearly tilted
    rate function, instead, remains strictly convex in the whole domain,
    $(I\circ F^{-1})''>0$. \label{fig:stretched-exponential-convexity}}
\end{figure}

This can be seen in figure~\ref{fig:stretched-exponential-pdf}: We
employ the numerical scheme by
Chernykh-Stepanov~\cite{chernykh-stepanov:2001,
  grafke-grauer-schaefer:2013} to compute the instanton starting at
$\varphi^*(t_0)=0$ and ending at $\varphi^*(t_1)=z$. The iterative algorithm
converges towards the minimizer of the action, and once converged, we
can estimate the probability of reaching $z$ by $p(z) \approx
\exp(-\eps^{-1} S(\varphi^*))$. As expected, though, computing the 
instanton fails beyond the inflection points at $z\approx \pm 1.26$, where
the tails become stretched exponentials (light blue dots): No
choice of $\lambda$ leads to endpoints $z$ of the instanton beyond
these, as the linear tilt diverges and the CGF is undefined. Instead,
we need to choose a non-linear tilt, such as
\begin{equation}
  F(z) = \sign(z) \log \left| z \right|  \,, \quad z\in\RR\setminus\{0\}\,,
\end{equation}
for which even in the tails the reparametrized expectation
$\EE[\exp(\eps^{-1}\lambda F(X^\eps_{t_1}))]<\infty$ remains bounded.
Consequently, the derivative of the CGF $G_F(\lambda)$ is a bijection,
so that every value of $\lambda$ has a corresponding $z$. This map is
explicitly given by
\begin{equation}
  \lambda(z) = \lambda \circ F^{-1}\left(x\right) = e^{4x} \left(4 + e^{3x}  \right)/\left(1 + e^{3x} \right)^2
\end{equation}
(for $z>0$, and negative for $z<0$).

With this choice, the instanton prediction for the stationary PDF is
almost exact far into the heavy tails (dark blue dots vs black solid in
figure~\ref{fig:stretched-exponential-pdf}). Here, we again employ the
iterative instanton computation, but are solving the instanton
equations with the boundary condition~(\ref{eq:bnd-cond-tilted})
instead. The underlying reason for convergence is that the
reparametrization with $F$ convexifies the rate function, i.e.~creating 
supporting lines with slopes $\lambda$ for all the domain of $I \circ F^{-1}$. 
As shown in figure~\ref{fig:stretched-exponential-convexity}, while the second
derivative of the rate function becomes negative beyond the inflection
points, the second derivative of the nonlinearly tilted rate function
remains positive throughout.

For the numerics in this example, we chose $\alpha=1$, with $N_t= 10^3$
timesteps, and a time interval of $T= 6$.

\subsection{Powerlaw distribution}
\label{sec:power-law-distr}

Even heavier tails are given by power law distributions,
\begin{equation}
  \label{eq:powerlaw}
  p(z) \sim |z|^{-\beta}\,,
\end{equation}
which are associated with a multitude of phenomena in
wide areas of science, in part due to their connection to scale
invariance, self-similarity, universality classes and criticality in phase
transitions. Here, we construct a simple SDE in $n=1$ dimensions which
has a powerlaw invariant density. Consider
\begin{equation}
  \label{eq:powerlaw-sde}
  dX_t^\eps = -\frac{\beta X_t^\eps}{1+(X_t^\eps)^2}\,dt + \sqrt{2\eps}\,dW_t\,.
\end{equation}
It can easily be shown that the invariant density for the
process~(\ref{eq:powerlaw-sde}) is given by
\begin{equation}
  \label{eq:powerlaw-pdf}
  \rho_\infty(z) = Z^{-1} (1+z^2)^{-\beta/2\eps}\,,
\end{equation}
where $Z$ is a normalization constant. For $z\gg1$ and $\eps=1$, this
takes the limiting form~(\ref{eq:powerlaw}), but is regularized for
small $z$. Again, we are interested in computing tail probabilities
for this toy model, by computing the instanton $\varphi^*$ realizing
large values of $z$, which yields the respective probability by
evaluating the corresponding action.

\begin{figure}
  \begin{center}
    \includegraphics[width=246pt]{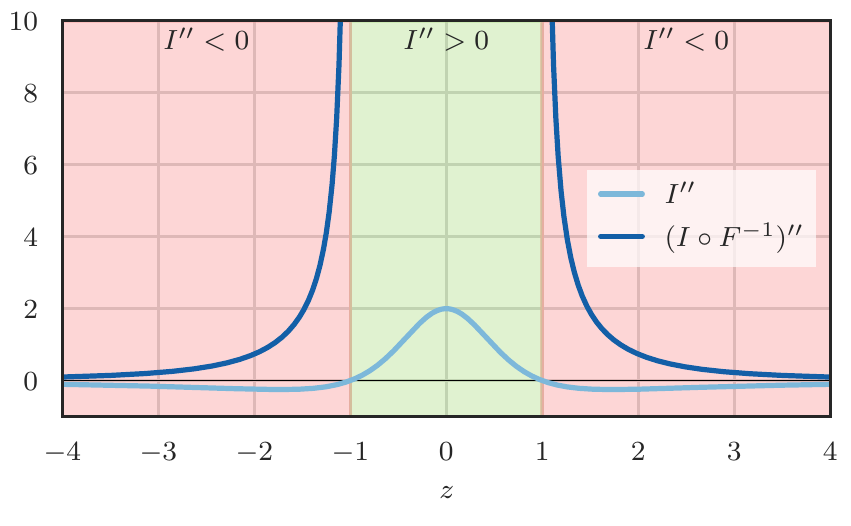}
  \end{center}
  \caption{Convexity condition for the powerlaw test case: The second
    derivative of the rate function is negative in the tails beyond
    the inflection points at $z= \pm 1$ (light blue). The nonlinearly tilted
    rate function is strictly convex in this region instead (dark
    blue). \label{fig:powerlaw-convexity}}
\end{figure}

\begin{figure}
  \begin{center}
    \includegraphics[width=246pt]{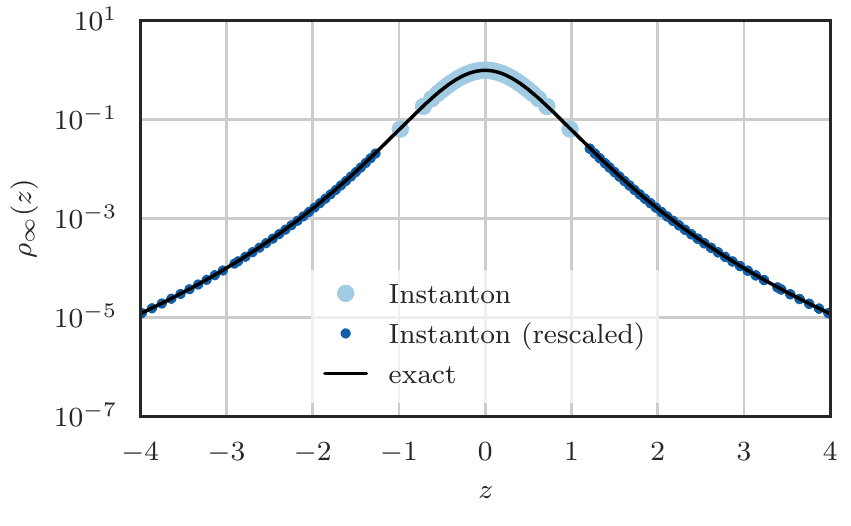}
  \end{center}
  \caption{Stationary density with powerlaw tails ($\beta=2,
    \eps=0.25$). Naively computing the instanton for tail events fails
    beyond $z=1$ (light blue). The nonlinear tilt $F(z) = \sign(z)
    \log\log |z|$ yields probabilities of events with $z\gg1$ within
    the non-convex powerlaw tail (dark blue) in good agreement with
    the theoretical result (black solid). \label{fig:powerlaw-pdf}}
\end{figure}

As in section~\ref{sec:stretch-expon-distr}, the LDT rate function,
given here by
\begin{equation}
  \label{eq:powerlaw-ratefunction}
  I(z) = \tfrac12 \beta \log(1+z^2)\,,
\end{equation}
does not admit supporting lines in the tails, and consequently its LF transform
is undefined (case~\ref{Case1} as well). This is reflected in the fact that the moment generating function
\begin{equation}
  \EE \exp(\eps^{-1}\lambda z) = \int_\RR \exp(\eps^{-1}\lambda z) (1+z^2)^{-\beta/2\eps}\,dz,
\end{equation}
diverges. We can convexify the rate
function~(\ref{eq:powerlaw-ratefunction}) by reparametrizing via
\begin{equation}
  F(z) = \sign(z) \log\log|z|\,,\quad z\in \RR\setminus [-1,1]\,,
\end{equation}
which is an even more drastic tail compression than needed for the
stretched exponential. Note that here we only convexify in the tails,
$|z|>1$, where the problem occurs, and do not attempt to find a global
map. Indeed, for this choice of $F$, the reparametrized rate function
$I\circ F^{-1}(z)$ is convex in the tails, as shown in
figure~\ref{fig:powerlaw-convexity}: Its second derivative remains
positive in the tails, which is not true for the original rate
function $I(z)$. We can explicitly map $\lambda$ to $z$ via
\begin{equation}
  \lambda(z) = \lambda \circ F^{-1}\left(x\right) = \beta e^{\left( 2 e^x + x\right) } / \left( 1+ e^{2 e^x}\right)
\end{equation}
(for $z>1$ and negative for $z<-1$). Numerically, as before, we solve
the optimization problem posed by the instanton equations with tilted
boundary conditions~(\ref{eq:bnd-cond-tilted}) to obtain instantons
$\varphi^*$ for events with large $z$. The corresponding action,
$S(\varphi^*)$, yields the tail probability. This computation is shown
in figure~\ref{fig:powerlaw-pdf}: The naive instanton computation
(light blue) leads to numerically diverging results in the tail
region, $|z|>1$, which are captured accurately by the reparametrized
instanton (dark blue). Parameters are $\beta=2$, $\eps=0.25$,
$N_t= 10^3$, and $T= 10$.

\subsection{Banana potential}
\label{sec:banana-potential}

\begin{figure}
	\begin{center}
		\includegraphics[width=246pt]{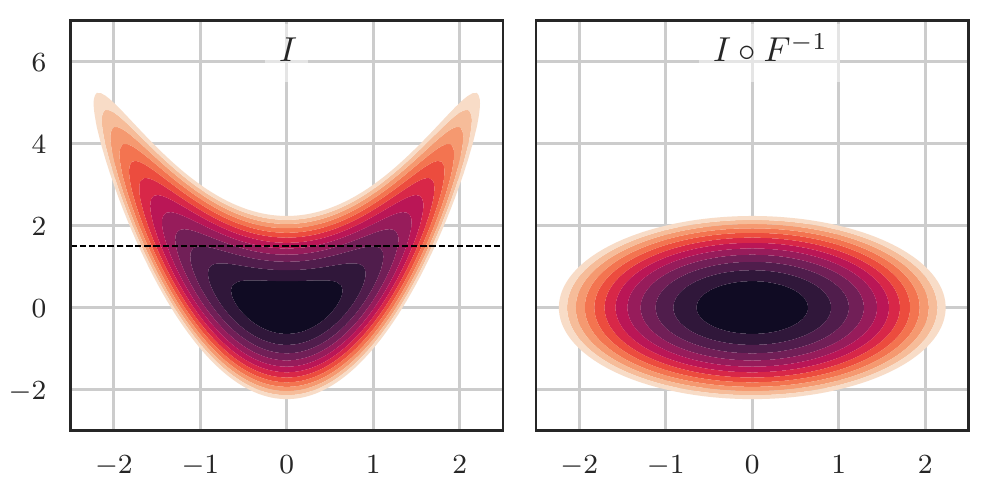}
	\end{center}
	\caption{Left: Contour plot of the rate function
      (\ref{eq:banana_rate_fct}) shows a banana-alike valley
      surrounding a non-convex plateau. The black dashed line
      represents the position of the marginal shown in
      figure~\ref{fig:Banana_SlicePDF}. Right: The rate function
      composed with the inverse of the nonlinear observable
      (\ref{eq:F_bananaEx}) deforms the landscape so that the rate
      function becomes strictly convex.  \label{fig:banana-potential}}
\end{figure}

In higher dimensions, non-convexity can manifest in more subtle ways
than in 1D. Consider for example the 2D system,
\begin{equation}
\label{eq:bananaEX_sde} 
dX^\eps_t =  b(X^\eps_t) \,dt + \sqrt{2\eps}\,dW_t\, ,
\end{equation}  
where, 
\begin{equation}
b(x)  =  - 2
\begin{bmatrix} 
  x_1 \left( 1 - 2  \ \left( x_2 - x_1^2 \right) \right) \\
  x_2  - x_1^2 
\end{bmatrix}. 
\end{equation}
This system is a gradient flow for the potential $U(x) = x_1^2 +
(x_2-x_1^2)^2$, so that again we have that the rate function for the
stationary distribution is equivalent to this potential, $I(z) =
U(z)$, i.e.
\begin{equation}
\label{eq:banana_rate_fct}
I\left( z \right) = z_1^2 + \left( z_2 - z_1^2 \right)^,.
\end{equation}
The system has a unique stable fixed point at the origin, which is the
deepest point of a banana-shaped valley (the set of points ${
  \left\lbrace \left( x_1, \ x_2\right) \ | \ x_2 = x_1^2
  \right\rbrace }$) of the potential, as can be seen in
figure~\ref{fig:banana-potential} (left). The rate function
$I\left(z\right)$ (\ref{eq:banana_rate_fct}) does not admit supporting
hyperplanes (\ref{eq:supp_hyp}) at the region $\left\lbrace z = \left(
a, \ b \right) | \ b > a^2 \right\rbrace$, leading to no tilt
variables $\lambda\in\mathbb R^2$ to reach an outcome $z$ within that
region.

Unlike the previous examples, the challenge in this case is therefore
not the far tails of the stationary density, but actually probing the
core of the distribution. The non-convexity of the rate function of
the previous examples amounted to the divergence of its LF transform,
the CGF (case~\ref{Case1}), while here the non-convexity leads to the
non-differentiability of the CGF (case~\ref{Case2}).

To fix this non-differentiability of the CGF $G(\lambda)$, we propose
a nonlinear reparametrization that satisfies the criteria of
(\ref{sec:prop-chos-observ}). Consider
\begin{equation}
\label{eq:F_bananaEx}
F \left(z \right)  =  
\begin{bmatrix} 
z_1 \\
z_2  - z_1^2 
\end{bmatrix}. 
\end{equation}
This reparametrization ``straightens the banana'', i.e.~it deforms the
space of outcomes such that the rate function becomes strictly convex,
as figure \ref{fig:banana-potential} (right) shows.
Using this reparametrization produces a continuous and differentiable CGF of 
the observable $F$, resulting from the LF transform of $I \circ F^{-1}\left(y \right)$,
\begin{equation}
\begin{split}
G_F\left( \lambda \right) & = \sup\limits_{F^{-1}\left(y \right) \in \RR^2}
\left(\left\langle \lambda ,y \right\rangle - I\circ F^{-1}\left(y \right)\right), \\
& = \frac{1}{4 } \, \left( \lambda_1^2 + \lambda_2^2\right), 
\end{split}
\end{equation}
which allows Lagrange multipliers to reach any outcome $z$, in
particular ones within the nonconvex region.

\begin{figure}
	\begin{center}
		\includegraphics[width=246pt]{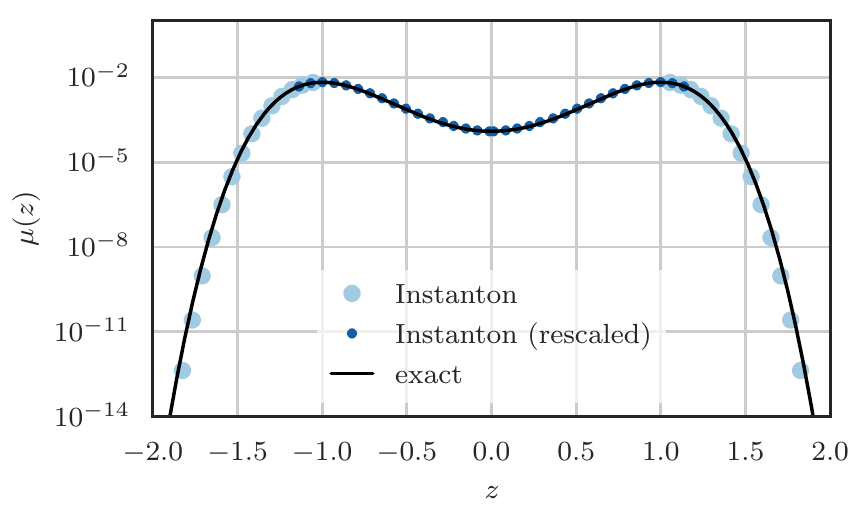}
	\end{center}
	\caption{The marginal distribution $\mu(z) =
      \rho^\infty(z,\tfrac32)$ (as denoted by the blacked dashed line
      in figure~\ref{fig:banana-potential} (left)).  Instantons with
      linear tilt (light blue) fail to reach the region $-1<z<1$
      without supporting hyperplanes of the rate function
      $I\left(z\right)$~(\ref{eq:banana_rate_fct}). Performing a
      reparametrization using the observable (\ref{eq:F_bananaEx})
      produces an strictly convex effective rate function ${I \circ
        F^{-1} \left(x\right)}$ that admits supporting hyperplanes
      everywhere. The corresponding instanton actions successfully
      capture the non-convex region (dark blue).
		 \label{fig:Banana_SlicePDF}}
\end{figure}

As numerical experiment, we choose to look at the marginal stationary
distribution $\mu$ in $z_1$ direction for a fixed value of
$z_2=\frac32$, i.e. $\mu(z) = \rho^\infty(z, \tfrac32)$. Since at any
fixed value $z_2>0$ we cut through the non-convex region $z_2>z_1^2$,
the marginal distribution $\mu(z)$ looks like a double-well
potential. We stress, though, that the whole system indeed has only a
single fixed point. We then solve the optimization problem posed by
the instanton equations with linear tilt, and compare to the
minimization problem with nonlinear tilt. As shown in
figure~\ref{fig:Banana_SlicePDF}, the linearly tilted instanton
computation produces acceptable results in the tails of the
probability density (light blue dots), it fails to converge within the
non-convex region $-1<z<1$: Since in that region there are no
supporting planes of the rate function (\ref{eq:banana_rate_fct}),
there is no $\lambda\in\mathbb R^2$ corresponding to the slope of the
supporting plane at that $z$, and consequently no tilt exists to
produce the desired outcome $z$.

For the reparametrized observable~(\ref{eq:F_bananaEx}), on the other
hand, the effective rate function $I\circ F^{-1}(x)$ is convexified
and admits supporting planes at every $z$. Indeed, as demonstrated in
figure~\ref{fig:Banana_SlicePDF}, the reparametrized optimization
problem leads instanton trajectories reaching outcomes (shown as dark
blue dots) within the non-convex region $-1<z<1$.

As a final remark, convexifying the rate function by the above method,
even though it guarantees the existence of a tilt for every outcome,
might nevertheless lead to numerical convergence issues. For example,
in regions where the original rate function was convex, the rescaled
optimization problem might be harder to solve, or necessitate more
iterations or smaller time-steps. Similarly, even in the convexified
region, the problem might become ill-posed, for example for $z_1\ll 1$
and $z_2>2$, where the observable is approximately linear.
 
\subsection{Nonlinear Schr\"odinger equation}
\label{sec:nonl-schr-equat}

\begin{figure*}
  \includegraphics[width=246pt]{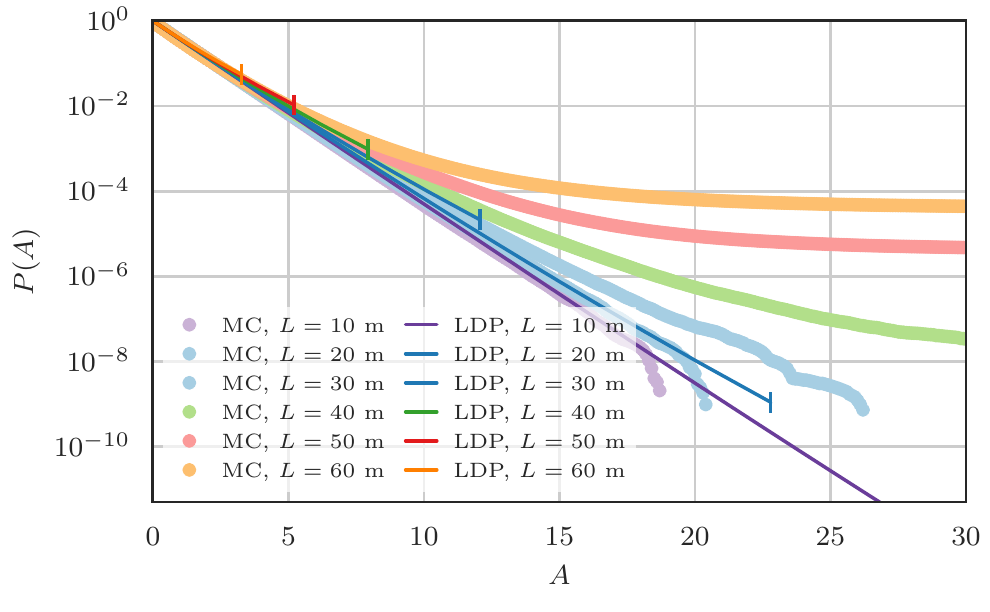}
  \includegraphics[width=246pt]{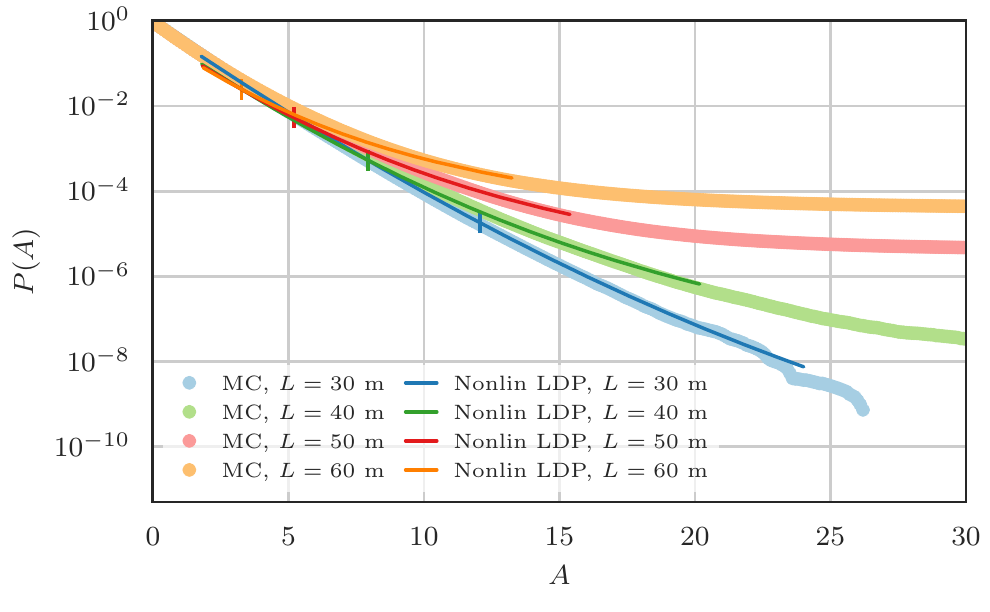}
  \caption{Left: PDF of optical power $A$ at the end of the
    fiber. Compared are Monte-Carlo simulations (MC, light color) with
    instanton prediction (dark color) for different fiber lengths,
    $L\in\{10\ \text{m},\dots,60\ \text{m}\}$. The tails become fatter
    with increasing $L$. A vertical marker is inserted at the maximal
    power achievable with the naive instanton method, highlighting how
    fat tails prevent useful instanton predictions. Right: The same,
    but for the instanton with nonlinear tilt,
    equation~(\ref{eq:NLS-tilt-function}). The LDP computation now
    reaches far into the fat tails. The vertical markers of the naive
    instanton are copied over here for comparison, highlighting the
    increased tail reach (e.g.~by more than a factor 3 for
    $L=60\ \text{m}$). Only lengths $L>30\ \text{m}$ are
    shown. \label{fig:NLS-tilt}}
\end{figure*}

As a practical example for our proposed method, we consider the
formation of extreme events in nonlinear wave
equations~\cite{zakharov:1968, osborne-onorato-serio:2000,
  mori-onorato-janssen-etal:2007, onorato-proment-el-etal:2016}. In
the field of nonlinear optics and photonics it has been established
that heavy tailed statistics frequently
occur~\cite{akhmediev-dudley-solli-etal:2013}. Physical mechanisms
such as soliton formation~\cite{kibler-fatome-finot-etal:2010,
  tikan-billet-el-etal:2017} and nonlinear
amplification~\cite{onorato-osborne-serio-etal:2005} are responsible
for the emergence of extreme power spikes out of incoherent, Gaussian
initial conditions, and have been subject to investigation by a
multitude of rare event algorithms~\cite{farazmand-sapsis:2017,
  dematteis-grafke-vanden-eijnden:2019}. 

Here, we consider the one-dimensional propagation of an optical pulse
along a fiber, described by the nonlinear Schr\"odinger equation (NLS)
\begin{equation}
  \label{eq:NLS}
  -i \partial_x \psi = \tfrac 12 \partial_t^2 \psi + |\psi|^2\psi\,,\quad \psi(x\!=\!0,t) = \psi_0(t),
\end{equation}
for a complex wave envelope $\psi: [0,L]\times[0,T] \to \CC$. Boundary
conditions are given at location $x=0$ at the beginning of the fiber
for all times $t\in[0,T]$, and the output is measured at the end of
the fiber at $x=L$. The input signal is considered random, with a
Gaussian distribution of known energy spectrum. Specifically, we are
mimicking an experimental setup such as~\cite{tikan:2018} of a
partially coherent light source, where the input signal is designed as
a Gaussian shape in frequency space with covariance
\begin{equation}
  \label{eq:NLS-chi}
  \chi_n \sim \exp(-\tfrac12 \omega_n^2/\Delta\nu^2)\,,\quad |n|<N\,,
\end{equation}
with spectral bandwidth $1/\Delta \nu$ and truncation frequency
$\omega_N$, so that the input signal is given by
\begin{equation}
  \label{eq:source-signal}
  \psi_0(t) = \sum_{n=-N}^N e^{i \omega t} \sqrt{\chi_n} \xi_n\,,
\end{equation}
where $\xi_n$ are i.i.d.~mean zero, unit variance complex Gaussian
random variables.

For this setup, we are interested in the probability of measuring
large spikes in the optical power $A(x,t)=|\psi(x,t)|^2$ at the fiber
end, $x=L$. Within the presented instanton formalism, this can be
achieved by tilting the distribution of initial conditions towards a
high-power outcome at the fiber end, and estimating the tail
probability by its most likely (``instantonic'') realization. The
corresponding LDP is given by
\begin{equation}
  \label{eq:NLS-p}
  p(z) = P[A(L,T/2)\ge z] \asymp \exp(-I(z))\,,
\end{equation}
for a power spike of size $z$ taken arbitrarily at the center of the
temporal domain, $t=T/2$. Due to the Gaussianity of the initial
conditions, the rate function $I(z)$ simply 
is~\cite{dematteis-grafke-vanden-eijnden:2019}
\begin{equation}
  \label{eq:NLS-I-tilted}
  I(z) = \inf_{\xi\in\CC^{2N+1}} \left(\tfrac12 |\xi|^2 - \lambda(z) |\psi(L,T/2)|\,\right)\,.
\end{equation}
Here, $\xi$ determines the source signal $\psi_0(t)$
through~(\ref{eq:source-signal}), while, $\lambda(z)$ can be
interpreted as a Lagrange multiplier enforcing the power constraint
$|\psi|^2=z$ at the end of the fiber. Equation~(\ref{eq:NLS-I-tilted})
is therefore simply saying that the rate function is given by the most
likely random configuration $\xi$ that determines a source signal with
high power output. Note that, similar to the examples above, the tilt
in~(\ref{eq:NLS-I-tilted}) is linear, and we can therefore expect the
expectation
\begin{equation}
  \EE \exp(-\lambda |\psi(L,T/2)|)
\end{equation}
over light source signals to diverge for fiber lengths $L$ long enough for
solitons to emerge and for the tails of the power distribution to
become fat.

Since the probability of high power output signals at the fiber end is
not known analytically, the only option we have to get comparison data
is to perform Monte-Carlo (MC) simulations to sample the power
distribution. To this end, we simulate the evolution of a wave packet
along the fiber with a random input signals with energy
spectrum~(\ref{eq:NLS-chi}) by numerically integrating
equation~(\ref{eq:NLS}). This equation is non-dimensionalized,
with $x$, $t$ and $\psi$ normalized by characteristic 
parameters $\mathcal{L}_0$, $\mathcal{T}_0$ and $\mathcal{P}_0$ 
respectively, such that
\begin{equation}
  x = \tilde{x} / \mathcal{L}_0, \qquad t = 
  \tilde{T} / \mathcal{T}_0, \qquad \psi = \tilde{\psi} / \sqrt{\mathcal{P}_0},
\end{equation}
where $\tilde{x}, \, \tilde{t}$ and $\tilde{\psi}$ are the
corresponding dimensional variables. These parameters
$\mathcal{L}_0$, $\mathcal{T}_0$ and $\mathcal{P}_0$ also determine
the dispersion $\beta_2$ and nonlinearity $\gamma$ properties of the
optical fiber via
\begin{equation}
  \beta_2 =   \mathcal{T}_0^2 / \mathcal{L}_0, \qquad  \gamma = 1 / \left( 
  \mathcal{L}_0 \, \mathcal{P}_0\right)\,.
\end{equation}
We chose these parameters according to the experimental setup in
\cite{tikan:2018}, where they are given as
\begin{equation}
  \mathcal{L}_0 = 160.3  \text{ m},  \ \mathcal{T}_0 = 
  1.8778\text{ ps}, \ \mathcal{P}_0 = 2.6 \text{ W}\,.
\end{equation}
Therefore, the optical fiber has dispersion parameter $\beta_2 = 0.022
\text{ ps$^2$/m}$ and nonlinearity constant $\gamma =
0.0024\ (\text{Wm})^{-1}$. The spectral bandwidth $1/\Delta \nu$ is
taken to be ($\Delta\nu = 0.5 \ \textrm{THz}$). 
We pick fiber lengths between
$10\ \text{m}$ and $60\ \text{m}$, periodic boundary conditions in
time treated pseudo-spectrally, and integrate with a second-order
Runge-Kutta exponential time differencing method
(ETDRK2)~\cite{du-zhu:2005} in the spatial variable. The
discretization is $\Delta x = 6.24 \times 10^{-3}, \ \Delta t = 1.3
\times 10^{-2}, T=106$, and frequency cut-off $N = 45$. As expected, 
and shown in figure~\ref{fig:NLS-tilt}, the tails of the PDF of optical power
become heavier with increasing fiber length $L$. Its samples number is $10^6$, 
where the property of $A$ being statistically homogeneous in time is used to 
improve the statistics.

To compare these brute-force sampling estimates to the instanton
prediction, we have to solve the optimization
problem~(\ref{eq:NLS-I-tilted}). This is done by defining the cost
functional
\begin{equation}
  E(\xi) = \tfrac12|\xi|^2 - \lambda |\psi(L,T/2)|
\end{equation}
for a given $\lambda$ and performing gradient descent, where the
gradient is given by
\begin{equation}
  dE/d\xi = \xi - \lambda J(L,T/2) d|\psi(L,T/2)|/d\psi\,,
\end{equation}
with Jacobian $J(x,t) = d\psi(x,t)/d\xi$. This gradient can be
evaluated by simultaneously integrating the NLS
equation~(\ref{eq:NLS}) and the evolution equation of the Jacobian,
\begin{equation}
  \label{eq:Jacobian}
  \partial_x J = i\left(\tfrac12 \partial_t^2 J + \psi^2 \bar J + 2 |\psi|^2 J\right),
\end{equation}
(where $\bar a$ is the complex conjugate of $a\in\CC$). The iterative
gradient descent algorithm yields the optimal choice $\xi^*$ that will
lead to the desired outcome of the final power exceeding the power
threshold $z$. As can be seen in figure~\ref{fig:NLS-tilt} (left), the
corresponding prediction for the probability, $\exp(-I(z))$ (from
equation~(\ref{eq:NLS-p})) correctly describes the tail decay of high
power events at the fiber end, but crucially \emph{only as long as the
  the rate function admits supporting lines, i.e.~remains convex}.
Therefore, the instanton prediction is basically useless for optical
fibers longer than $L=30\ \text{m}$. As a side note, the gradient
computation could instead be performed in the adjoint formalism,
leading to two coupled forward-backward equations similar in spirit to
the instanton equations~(\ref{eq:varphi-p}), but identically yielding
meaningful results only in the convex region of the rate function.

Now, applying the idea from above, we can instead \emph{nonlinearly
  tilt} the probability distribution of input signals towards high
power outcomes. For this, we choose instead the nonlinearly tilted
rate function
\begin{equation}
  \label{eq:NLS-I-nonlin-tilted}
  I(x) = \inf_{\xi\in\CC^{2N+1}}\left( \tfrac12|\xi|^2 - \lambda(x) F(\psi(L,T/2)\right)
\end{equation}
with
\begin{equation}
  \label{eq:NLS-tilt-function}
  F(z) = \log\log |z|\,,\quad |z|>1\,.
\end{equation}
For this tilt, the cost functional becomes
\begin{equation}
  E(\xi) = \tfrac12|\xi|^2 - \lambda F(\psi(L,T/2))
\end{equation}
and the gradient is
\begin{equation}
  dE/d\xi = \xi - \lambda J(L,T/2) dF(\psi(L,T/2))/d\psi\,,
\end{equation}
instead. The results of this are shown in figure~\ref{fig:NLS-tilt}
(right) for the four longest fiber lengths of $30-60\ \text{m}$ in
$10\ \text{m}$ increments, where the tails are fattest. In the revised
formalism (dark color), the nonlinearly tilted instanton prediction is
able to reach far into the stretched tail and gives the right order of
magnitude for the probability of power spikes obtained from sampling
(light color). The end of the region of convergence for the naive
instanton is shown for comparison (vertical markers).

Note that due to the choice of reparametrization $F$
in~(\ref{eq:NLS-tilt-function}), the nonlinearly tilted instanton
prediction is restricted to the region of normalized power
$|\psi|^2>1$, but of course this is exactly the tail region that we
care about.

\section{Conclusion}
\label{sec:conclusion}

Estimating the probability of tail events can efficiently be done via
large deviation theory and instanton calculus, which transforms an
inefficient sampling problem into a deterministic optimization
problem. Unfortunately, for systems with heavy tails, or more
generally non-convex rate functions, standard mechanisms of
exponentially tilting the measure, or numerically solving the
optimization problem, fail. The reason is the absence of a bijective
map between Lagrange multiplier (tilting parameter) and desired
outcome, caused by the breakdown of their Lagrange duality, or
equivalently by the non-convexity of the rate function.

We put forward the idea of a nonlinear tilt that reparametrizes the
output space, effectively convexifying the rate function of the observed 
probability distribution. We discuss the necessary conditions required 
for this reparametrization to yield a unique outcome variable and 
ensure a bijective mapping between tilt and outcome: It needs to be a
diffeomorphism chosen such that its composition with the rate function
is strictly convex.  Note further that the reparametrization can be
chosen locally, i.e. the conditions on the nonlinear observable need
only apply in a subdomain of the events of interest.

Finding such nonlinear observable can be subtle, especially when the
system is highly nonlinear, influencing the rate function
landscape. However, drawing inspiration from toy problems with
stretched exponential and algebraic tails, which can be treated
analytically, yields candidate reparametrizations for physically
relevant problems. We show the applicability to real-world problems by
demonstrating how instantons determine the probability in extreme
optical power events in a fiber optical cable, where solitons lead to
a heavy-tailed power distribution at the fiber end.

\section{Acknowledgement}

The authors thank Eric Vanden-Eijnden for helpful discussions and
Giovanni Dematteis for help with the source code for the fiber optics
example. MA acknowledges the PhD funding received from UKSACB. TG
acknowledges the support received from the EPSRC projects EP/T011866/1
and EP/V013319/1.

\addcontentsline{toc}{section}{Bibliography}
\bibliographystyle{apsrev4-1}
\bibliography{bib}

\end{document}